\documentstyle[12pt] {article}
\begin {document}
\begin{center}
\vskip 1.5 truecm
{\bf t-DEPENDENCES OF VECTOR MESON DIFFRACTIVE PRODUCTION
IN $ep$ COLLISIONS} \\

\vspace{.5cm}
M.G.Ryskin, Yu.M.Shabelski and A.G.Shuvaev \\
\vspace{.5cm}
Petersburg Nuclear Physics Institute, \\
Gatchina, St.Petersburg 188350 Russia \\

\end{center}
\vspace{1cm}
\begin{abstract}
We calculate $p_T$-dependences of diffractive vector meson photo- and
electroproduction in electron-proton collisions with and without proton
dissociation at small momenta transfers. The calculated slopes are in
good agreement with the data.

\end{abstract}
\vspace{3cm}

E-mail: $\;$ RYSKIN@THD.PNPI.SPB.RU  \\

E-mail: $\;$ SHABEL@VXDESY.DESY.DE \\

E-mail: $\;$ SHUVAEV@THD.PNPI.SPB.RU  \\
\newpage

Without any specified model we can consider diffractive production
of vector meson in high energy $ep$ collisions as following. A
virtual photon can turn with probability $\alpha_{em}$ = 1/137 into some
composite hadronic state (virtual vector meson, $V^*$) and some
constituent of this state interacts with a constituent of the target
proton. After that we obtain the real vector meson, $V$, without
(Fig. 1a) or with (Fig. 1b) target proton dissociation. The cross
section of "elastic" (without proton dissociation) vector meson photo-
or electroproduction with small transfer momentum to the proton has the
form
\begin{equation}
\frac{d\sigma (\gamma p \to Vp)}{dt}\; =\; F_p^2(t) F_V^2(t,Q^2)\;
|A(s,t,Q^2)|^2 \;,
\end{equation}
where
$F_p(t)$ and $F_V(t)$ are form factors of the proton and vector meson,
respectively, that account for the probabilities of their "elastic"
production and $A(s,t)$ is the amplitude of constituent interaction.
This expression is one of the results of additive quark model
\cite{AKNS}, however really it needs only the assumption that
pomerons and photons interact with hadrons by similar way
(pomeron-photon duality).

The similar cross section of "inelastic" (with proton dissociation)
vector meson production does not contain the proton form factor
\begin{equation}
\frac{d\sigma
(\gamma p \to VY)}{dt}\; =\; F_V^2(t,Q^2)\; |A(s,t,Q^2)|^2 \;.
\end{equation}

The proton form factor at moderate values of $t$ is parametrized usually
by dipole expression
\begin{equation}
F_p(t) = \frac{1}{\left ( 1 - \frac{t}{m^2}\right )^2 }
\end{equation}
with $m^2$ = 0.71 GeV$^2$.

Form factors of vector mesons are unknown experimen\-tally. It seems to
be reasonably to assume that they are similar to the form factors of
$\pi$ and $K$ mesons which have monopole form, so we use
\begin{equation}
F_V(t, Q^2) = \frac{1}{\left ( 1 + \frac{Q^2 + |t|}{M_V^2}\right ) }
\end{equation}
where $M_V$ is the vector meson mass. The argument $Q^2 + |t|$ gives the
possibility to describe simultaneously the case of photoproduction
($Q^2 = 0$) where the value $F_V$ coinside with a standard monopole
formfactor, and the case of high-$Q^2$ electroproduction where the
appearing factor $1/Q^4$ in Eqs. (1) and (2) corresponds to the
propagator of $V^*$ in Figs. 1a and 1b.

Amplitude $A(s,t,Q^2)$ at high energies can be taken in Regge form as
an effective one-pomeron (vacuum singularity) exchange
\begin{eqnarray}
A(s,t,Q^2)\,&=&\, A_0 \left(\frac{\mu^2}{Q^2+M_V^2}\right)^\gamma\,
\exp{(B_P t)} \nonumber \\
&=&\, A_0 \left(\frac{\mu^2}{Q^2+M_V^2}\right)^\gamma\,
\exp { \left (\alpha '_P t \ln \frac{s}{s_Q}\right ) } \;,
\end{eqnarray}
where the effective slope $\alpha '_P \approx 0.15$ \cite{Burq},
$s = W^2$ is the interaction energy square and $s_Q$ increases with $Q^2$.
This effect which reduce the total slope of $d\sigma /dt$ in Eqs. (1) and (2)
for the case of large photon virtuality $Q^2$ (and/or large mass of
produced vector meson) is the following. The transverse size of a
pomeron near the photon-vector meson vertex should be very small; the
virtuality and transverse momenta $k_T$ of gluons which form this part
of the pomeron are rather large, $k_T^2 \sim Q^2 + M_V^2 \gg \mu^2$
($\mu^2 \sim m_{\rho}^2$). On the other hand the value of $\alpha '_P$
is determined by the typical gluon transverse momenta,
$\alpha '_P \propto \alpha_s/k_T^2$, and the pomeron, build up by the
high $k_T$ gluons only, would have negligibly small value of $\alpha '$.

So one has to wait some time of evolution (some interval of rapidity
$y = \ln{1/x}$) before the initial small size component of the pomeron
gets into the normal equilibrium state with nonzero value of
$\alpha '_P \propto \alpha_s/\mu^2$.

Another consequence of the same effect is the factor
$\bigl(\mu^2/(Q^2+M_V^2)\bigr)^\gamma$ included in Eq (5). It reflects the
fact that the amplitude originated from the small size component of the
pomeron is less than the normal one. The suppression factor is controlled
by the anomalous dimension $\gamma$.

The precise formulae for this "hard" pomeron are based on the BFLK
equation for the non-forward QCD pomeron \cite{LL}. Unfortunately the
presented expressions are too complicated, therefore we will use the
simplified estimati\-ons of the effect.

According to the Lipatov $\log{k_T^2}$ diffusion, the variation of
$\ln{k_T^2}$ at each step of BFKL evolution is
$\Delta \ln{k_T^2} \approx 2$. To be more precise we have to write
$\Delta \ln{k_T^2} = 1/\gamma$, where the anomalous dimention $\gamma$,
which corresponds to the extremum of the BFLK solution, tends at high
energies to the value $\gamma = 1/2$.

By the same way the rapidity interval $\Delta y = \Delta \ln{1/x}$
needed for one step of evolution (diffusion) can be expressed in term
of the pomeron intercept $\lambda = \alpha_P(0) - 1$. Numerically for
BFLK pomeron $\Delta \ln{1/x} = 1/\lambda \approx 2$ and it means that
in order to diminish the initial large value of
$k_{T(in)}^2 \sim Q^2 + M_V^2$ down to the standard value for the soft
processes $k_T^2 \sim \mu^2$, which corresponds to the "equilibrium"
state of the pomeron with nonzero value
$\alpha '_P \approx$ 0.15 GeV$^{-2}$, one has to spent an interval
$\delta y \approx \frac{\Delta \ln{1/x}}{\Delta \ln{k_T^2}}
\ln{\frac{k_{T(in)}^2}{\mu^2}} = \frac{\gamma}{\lambda}
\ln{\frac{Q^2 + M_V^2}{\mu^2}} \approx
\ln{\frac{Q^2 + M_V^2}{\mu^2}}$\footnote{Note that the experimental
ratio of the typical values of anomalous dimention $\gamma$
($F_2(x,Q^2) \sim x^{-\lambda}(Q^2)^{\gamma}$) and intercept $\lambda$
is close to $\gamma /\lambda \approx 1$ within the HERA domain, where
$Q^2$-dependence is predicted. Recall that at large $W^2$ near the
extremum of BFKL $\gamma = 1/2$.}.

Thus we can write the slope of the pomeron amplitude of Eq. (5) in the
form
\begin{equation}
B_P = \alpha '_P \left ( \ln {\frac{1}{x}} -
\ln{\frac{Q^2 + M_V^2}{\mu^2}}\right ) =
\alpha '_P \ln {\frac{s\mu^2}{(Q^2 + M_V^2)^2}}
\end{equation}
with $\mu^2 = m_{\rho}^2$.

The experimental cross section $\frac{d\sigma (\gamma p \to Vp)}{dt}$
is usually parametrized as $e^{bt}$ or $e^{bt + ct^2}$. If we write
\begin{equation}
\frac{d\sigma (\gamma p \to Vp)}{dt}\; \propto\; e^{b_V(t) \cdot t} \;,
\end{equation}
we obtain in our parametrization for the case of "elastic" production,
using Eqs.  (1), (3) and (4)
\begin{equation}
b^{el}_V(t) = \frac{4}{m^2 + |t|} + \frac{2}{M_V^2  + Q^2 + |t|} +
2 B_P
\end{equation}
and for the case of "inelastic" production
\begin{equation}
b^{inel}_V(t) = \frac{2}{M_V^2 + Q^2 + |t|} +
2 B_P \;.
\end{equation}
In the case of hadron-hadron interactions the expression similar to Eq.
(8) for $b^{el}(t)$ describes quite good \cite{AKNS} the experimental
data at different $t$ for $\pi p$ and $pp$ scattering at $p_{lab}$ = 200
GeV/c.

Let us now compare the theoretical predictions, Eqs. (8) and (9) with
the experimental data, see Tables 1 and 2. In the cases where there are
many experimental points (say, for $\rho (0)$ photo- and
electroproduction) we present the data having the smallest experimental
errors.

The most accurate value of the slope is obtained for the case of
$\rho^0$ "elastic" photoproduction \cite{ZEUS1} where the error is only
about 2\% and the dependence of $b(t)$ on the $t$ region is evident. The
agreement with Eq. (8) is very good for so simple theoretical
estimations. The predicted value of $d\sigma (\gamma p \to \rho p)/dt$
normalized to the data with the help of parameter $A_0$ of Eq. (5) is
compared with experimental data of ref. \cite{ZEUS1} in Fig. 2a and one
can see that the agreement is quite good.

In the case of $\rho^0$ elastic electroproduction the values of the
slopes decrease significantly in agreement with our predictions.

The result for $\omega$ photoproduction \cite{ZEUS2} is of the order of
the $\rho^0$ case.

In the case of $\phi$ photoproduction the value of the slope was
obtained in the region 0.1 GeV$^2 < |t| <$ 0.5 GeV$^5$, so we present
the theoretical value of the slope at $t$ = 0.3 GeV$^2$. The slope
values here also decrease in the case of electroproduction.

The slopes decrease with the increase of the mass of produced vector
meson as one can see from Eqs. (4) and (6). This effect should be more
significant in the case of small-$Q^2$ interactions. One can see it on
the example of $\rho '$ electroproduction as well as of $J/\psi$
"elastic" photo- and electroproduction. In Fig. 2b we compare our
predictions for $d\sigma (\gamma p \to J/\psi p)/dt$ normalized to the
data with the experimental points of ref. \cite{H14} and one can see
that the predicted slope is really slightly too large as it is presented
in Table 1.

Now let us consider the slopes in the "inelastic" photo- and
electroproduc\-tion, i.e. with diffractive dissociation of the proton.
We compare our predictions given by Eq. (9) with the data in Table 2.
In all cases the calculated values of the slopes are in good agreement
with the data.

As a result we can clime that the simplest estimations of the slopes in
the processes of vector meson photo- and electroproduction are in
supprising agreement with all available experimental data.

The slightly different expression for differential cross sections of
vector meson photo- and electroproduction was presented in \cite{HKK}. 

This work is supported by Volkswagen Stiftung and, in part, by Russian
Fund of Fundamental Reswarch (95-2-03145), INTAS grand 93-0079 and NATO
grand OUTR. LG 971390.

\newpage
\begin{table}
\begin{center}
{\bf Table 1}
\end{center}
The comparison of predictions of Eq. (8) for the slopes of "elastically"
produced vector mesons, $b(0)$ (in GeV$^{-2}$), with the experimental
data.

\vskip 0.5 truecm
\begin{center}
\begin{tabular}{|c|c|c|c|c|} \hline
Reaction & $Q^2$, GeV$^2$ & $<\!W\!>$, GeV & $b(0)$, Eq. (8)
& $b(0)$ (exp) \\ \hline \hline

$\gamma p \to \rho p$ & 0 & 70 & 11.7 & $11.6 \pm 0.2$ \cite{ZEUS1} \\
\hline

$\gamma p \to \rho p$ & 10 & 80 & 6.9 & $7.8 \pm 1.0 \pm 0.7$ \cite{H11}
\\ \hline

$\gamma p \to \rho p$ & 20 & 80 & 6.4 & $5.7 \pm 1.3 \pm 0.7$ \cite{H11}
\\ \hline

$\gamma p \to \omega p$ & 0 & 70 & 11.6 & $10.0 \pm 1.2 \pm 1.3$
\cite{ZEUS2} \\ \hline

$\gamma p \to \phi p \;^1$
& 0 & 70 & 7.8 & $7.3 \pm 1.0 \pm 0.8$ \cite{ZEUS3} \\ \hline

$\gamma p \to \phi p$ & 10 & 100 & 7.0 & $5.2 \pm 1.6 \pm 1.0$ \cite{H12}
\\ \hline

$\gamma p \to \rho ' p$ & 10 & 80 & 6.8 & $5.5 \pm 1.9 \pm 0.5$
\cite{H13} \\ \hline

$\gamma p \to J/\psi p \;^2$ & 0 & 90 & 4.7 &
$4.0 \pm 0.2 \pm 0.2$ \cite{H14} \\ \hline

$\gamma p \to J/\psi p \;^3$ & 12 & 90 &
 4.1 & $3.8 \pm 1.2^{+2.0}_{-1.6}$ \cite{H11} \\ \hline

\end{tabular}
\end{center}
$^1$ At $<|t|>$ = 0.3 GeV$^2$ \\
$^2$ At $<|t|>$ = 0.5 GeV$^2$ \\
$^3$ At $<|t|>$ = 0.5 GeV$^2$ \\

\begin{center}
{\bf Table 2}
\end{center}
The comparison of predictions of Eq. (9) for the slopes of
"inelastically" produced vector mesons, $b(0)$ (in GeV$^{-2}$), with
the experimental data.

\vskip 0.5 truecm
\begin{center}
\begin{tabular}{|c|c|c|c|c|} \hline

Reaction & $Q^2$, GeV$^2$ & $<\!W\!>$, GeV & $b(0)$, Eq. (9) &
$b(0)$ (exp) \\ \hline \hline

$\gamma p \to \rho Y$ & 0 & 70 & 6.1 & $5.3 \pm 0.3 \pm 0.7$
\cite{ZEUS1} \\ \hline

$\gamma p \to \rho Y$ & 10 & 100 & 1.4 & $2.1 \pm 0.5 \pm 0.5$
\cite{H12} \\ \hline

$\gamma p \to J/\psi Y \;^1$ & 0 & 100 & 1.45
& $1.6 \pm 0.3 \pm 0.1$ \cite{H12} \\ \hline

\end{tabular}
\end{center}
$^1$ At $<|t|>$ = 0.5 GeV$^2$ \\

\end{table}

\newpage

\begin{center}
{\bf Figure captions}\\
\end{center}

Fig. 1. Diagrams for "elastic" (a) and "inelastic" (b) diffractive
vector meson photo- and electroproduction.

Fig. 2. Differential cross sections of $\rho^o$ (a) and $J/\psi$ (b)
elastic photopro\-duc\-tion. The predicted cross sections are normalized to
the data.


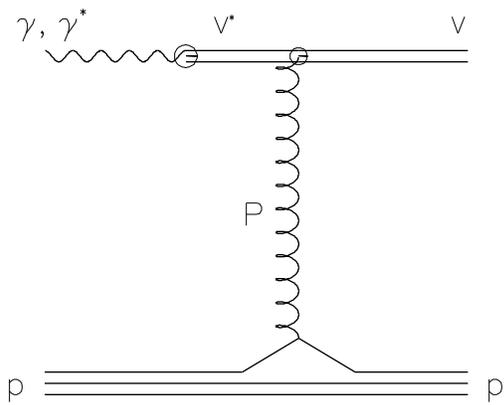
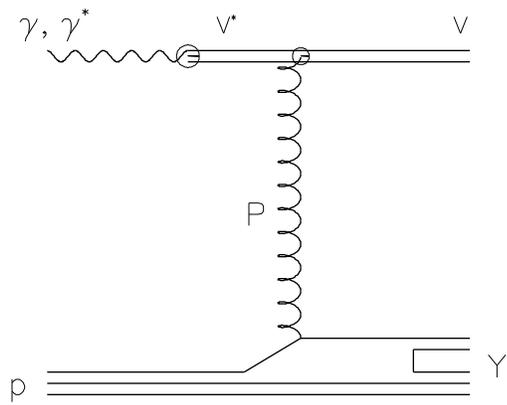
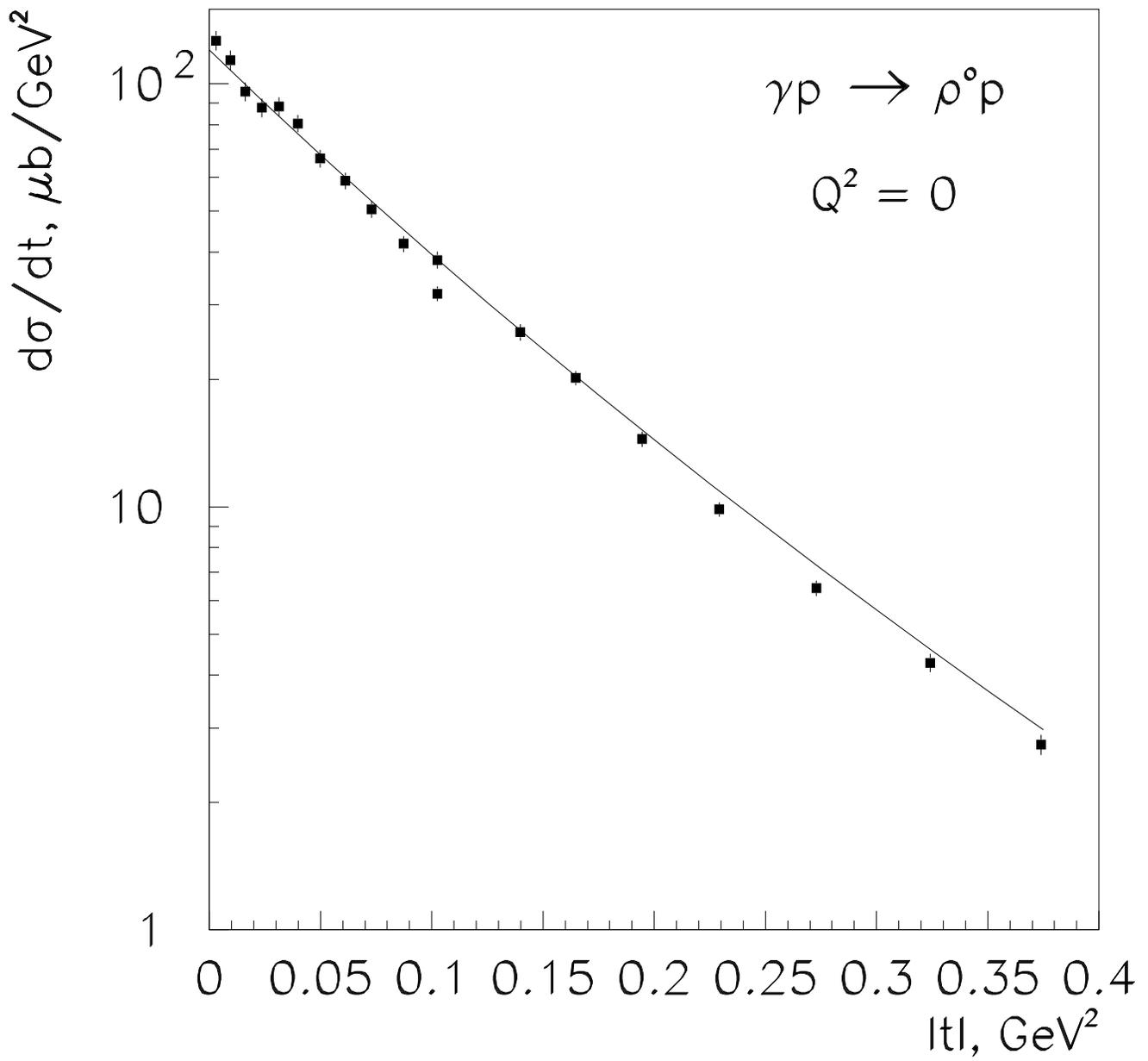
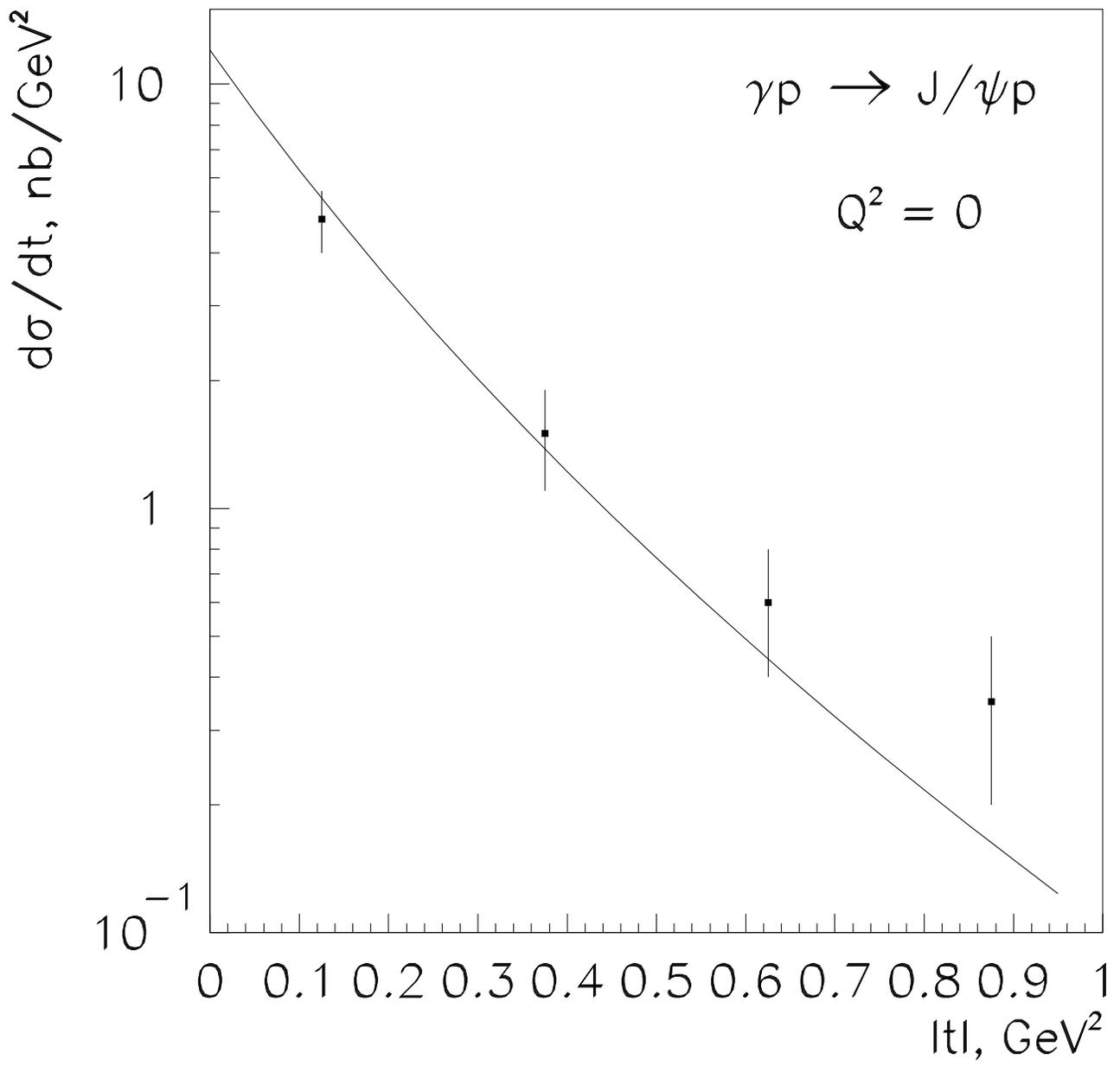
\end{document}